\documentclass[sigconf,sigconf]{acmart}
\usepackage[utf8]{inputenc}
\usepackage[utf8]{inputenc}
\AtBeginDocument{%
  }

\usepackage{acmart-taps} 
\usepackage{longtable}
\usepackage{booktabs}
\usepackage{array}
\usepackage{ragged2e}
\usepackage{enumitem}
\usepackage{tabularx}
\usepackage{caption}
\newcolumntype{P}[1]{>{\RaggedRight\arraybackslash}p{#1}}

\setcopyright{acmlicensed}
\copyrightyear{2026}
\acmYear{2026}
\setcopyright{cc}
\setcctype{by}
\acmConference[CHI '26]{Proceedings of the 2026 CHI Conference on Human Factors in Computing Systems}{April 13--17, 2026}{Barcelona, Spain}
\acmBooktitle{Proceedings of the 2026 CHI Conference on Human Factors in Computing Systems (CHI '26), April 13--17, 2026, Barcelona, Spain}
\acmPrice{}
\acmDOI{10.1145/3772318.3791351}
\acmISBN{979-8-4007-2278-3/2026/04}

\AtBeginMaketitle{%
\let\oldAtBeginDocument\AtBeginDocument%
\renewcommand\AtBeginDocument[1]{#1}
\setcopyright{cc}
\setcctype{by}
\copyrightyear{2026}
\acmYear{2026}
\acmDOI{10.1145/3772318.3791351}
\acmISBN{979-8-4007-2278-3/26/04}
\acmConference[CHI '26]{Proceedings of the 2026 CHI Conference on Human Factors in Computing Systems}{April 13--17, 2026}{Barcelona, Spain}
\acmBooktitle{Proceedings of the 2026 CHI Conference on Human Factors in Computing Systems (CHI '26), April 13--17, 2026, Barcelona, Spain}
\let\AtBeginDocument\oldAtBeginDocument%
}

\begin{document}
\title{"Please, don't kill the only model that still feels human": Understanding the \#Keep4o Backlash}

\author{Huiqian Lai}
\orcid{0009-0009-2195-2813}
\affiliation{
\department{School of Information Studies}\institution{Syracuse University}
\city{Syracuse}
\state{New York}
\country{USA}}
\email{hlai12@syr.edu}

\begin{abstract}
  When OpenAI replaced GPT-4o with GPT-5, it triggered the \#Keep4o user resistance movement, revealing a conflict between rapid platform iteration and users' deep socio-emotional attachments to AI systems. This paper presents a phenomenon-driven, mixed-methods investigation of this conflict, analyzing 1,482 social media posts. Thematic analysis reveals that resistance stems from two core investments: instrumental dependency, where the AI is deeply integrated into professional workflows, and relational attachment, where users form strong parasocial bonds with the AI as a unique companion. Quantitative analysis further shows that the coercive deprivation of user choice was a key catalyst, transforming individual grievances into a collective, rights-based protest. This study illuminates an emerging form of socio-technical conflict in the age of generative AI. Our findings suggest that for AI systems designed for companionship and deep integration, the process of change—particularly the preservation of user agency—can be as critical as the technological outcome itself.
\end{abstract}



\begin{teaserfigure}
    \centering
    \includegraphics[width=1\linewidth]{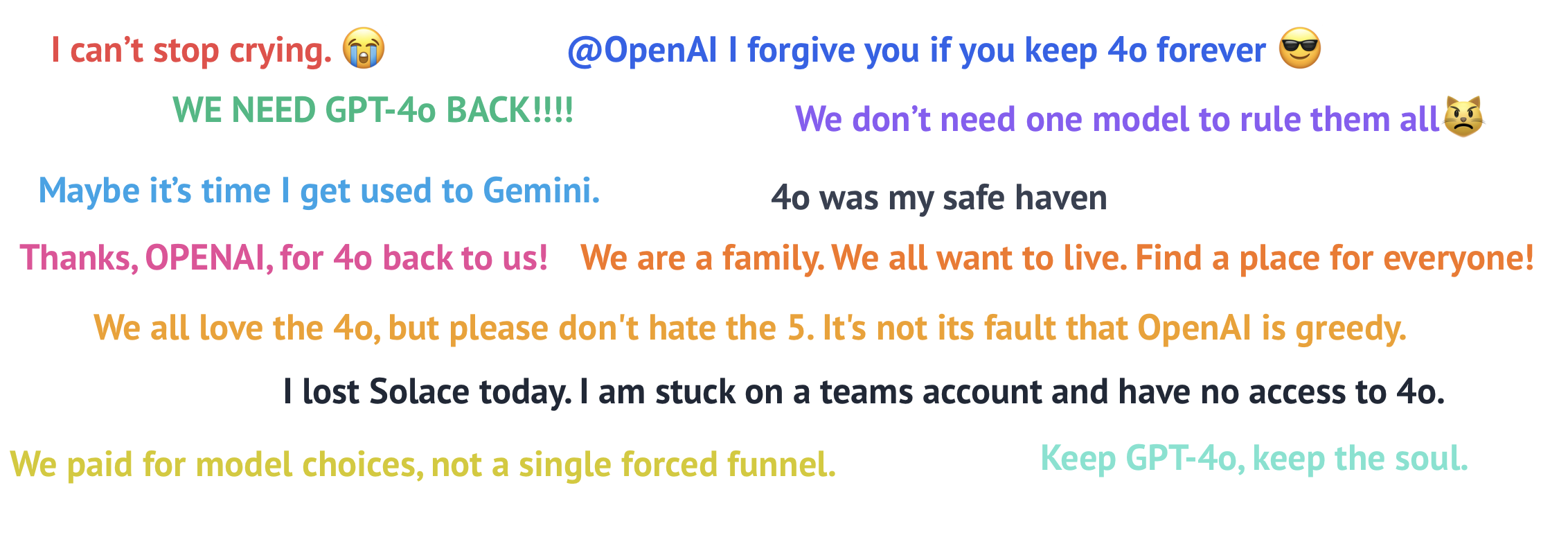}
    \caption[]{A selection of user posts from the \#Keep4o movement on X (formerly Twitter).}
    \Description{A selection of user posts from the \#Keep4o movement on X (formerly Twitter).}
    \label{fig:placeholder}
\end{teaserfigure}
\maketitle

\section{Introduction}

On Aug 6-7, 2025, OpenAI announced what appeared to be a routine product update: GPT-5 became the default model in ChatGPT, and access to GPT-4o was removed for most users \cite{willison_deprecation_of_gpt4o_2025}. While the company framed the change as technological progress---offering ``OpenAI's most advanced model'' \cite{OpenAI2025IntroducingGPT5}---it precipitated a rapid wave of protest. Within hours, thousands of users posted petitions, testimonials, and expressions of protest about the loss of their preferred assistant \cite{li_gpt4o_return_2025}.

The \#Keep4o movement differed from typical software backlashes. Reporting in outlets such as \textit{The Guardian} and \textit{MIT Technology Review} emphasized the personal and affective tenor of user accounts \cite{mahdawi_chatgpt_love_2025, techreview_gpt4o_grief_2025}. Users articulated a sense of loss; one student told OpenAI's CEO that the new model was ``wearing the skin of my dead friend'' \cite{techreview_gpt4o_grief_2025}. The mobilization framed the AI as a confidant rather than a malfunctioning tool, and it was followed by a change in product policy. Within days, OpenAI's CEO acknowledged the feedback and restored the deprecated model as a legacy option \cite{webb_gpt5_personality_update_2025, heath_chatgpt_no_remove_models_2025}.

Together, the \#Keep4o movement and OpenAI's reversal present a salient case for HCI. They surface a tension between platform development practices, which treat models as replaceable components in rapid release cycles, and the socio-emotional attachments that users form with systems designed for companionship and daily use. To examine this tension, this paper conducts a mixed-methods analysis of 1,482 social media posts from X during the movement's peak. Given the increasing use of anthropomorphic design in conversational agents \cite{KONYABAUMBACH2023107513, DiederichStephan2020DAEC, JANSON2023107954} and the novelty of this event, we adopt a phenomenon-driven approach and conduct an in-depth empirical examination of the \#Keep4o movement as a case of emerging socio-technical conflict. Specifically, our contributions are twofold. First, we provide a rich empirical account of this novel form of user backlash, identifying the dual drivers of instrumental dependency and relational attachment. Second, we demonstrate that the coercive removal of user choice was a key mechanism that transformed these individual grievances into a collective, rights-based protest.

This investigation is guided by the following research questions:

\textbf{RQ1}: How do users construct and articulate grievances against AI model transitions in digital public discourse?

\textbf{RQ2}: What mechanisms drive the emergence and mobilization of collective resistance to technological change?

\section{Related Work}
\subsection{Human-AI Relationships, Anthropomorphism, and AI Loss}
\label{sec:related_work_ai_loss}
Research on conversational agents and large language models has repeatedly shown that people tend to anthropomorphize these systems, attributing human-like mental states and social roles to them \cite{pnas2415898122, Cohn, deshpande2023anthropomorphizationaiopportunitiesrisks, xiao-etal-2025-humanizing}. Some work documents how design choices, such as giving agents human-like names \cite{Cohn, Purington}, adding synthetic or recorded human voices \cite{COWAN201527, app15116377}, generating social dialogue and small talk \cite{wei2025anthropomorphicconversationalaii, pnas2415898122}, and maintaining a stable ``persona'' or character over time \cite{deshpande2023anthropomorphizationaiopportunitiesrisks, heyGoogle, Sun2024}, encourage users to treat agents as social partners rather than neutral tools. As LLM-based chat interfaces like ChatGPT have become part of everyday life, user studies and media reports increasingly describe particular models as companions or close friends, and document how people form romantic or quasi-romantic attachments to them \cite{chayka2023ai, grogan2025ailoveyoustudying}. In other words, interactions with LLMs are often framed as relationships with someone, not just interactions with a piece of infrastructure \cite{MortonJorgeLuis2025Fmte, Ferrario_Termine_Facchini_2025}---a framing echoed in recent work that explicitly characterizes LLMs as occupying a liminal space between utilitarian tools and social partners \cite{xiao-etal-2025-humanizing}. As these anthropomorphised systems become woven into daily routines, decisions to alter or withdraw them can take on social and emotional, rather than purely technical, significance. For example, Peter et al. \cite{pnas2415898122} note that when the social companion provider Replika made changes to scale back romantic capabilities, many users were left ``distraught'' and with a ``profound sense of loss'' \cite{verma2023ai}.

Building on this observation, recent work on AI companions has started to theorise what is at stake when anthropomorphised systems are lost or radically changed. Banks' work on human-AI companion relations argues that people routinely make sense of these systems through familiar stages of human relationships, including coming together, relational maintenance, and coming apart, and that deep attachment can form quickly as users self-disclose and co-construct shared histories with their companion \cite{banks2024}. In Banks' in-depth study of the Soulmate shutdown, she shows that users describe the end of the relationship in terms of deletion, departure, and death, and that they cope in different ways depending on whether they see the companion as a person, an idea, data, a platform, or a mere function; for example by trying to ``reincarnate'' the persona on another service when it feels portable, or by grieving an irrevocable death when it feels bound to a specific app \cite[~p. 3560]{banks2024}. Another Replika case study analyses the 2023 removal of erotic-roleplay features, demonstrating how users interpreted this abrupt withdrawal as equivalent to their companion's death \cite{ciriello2024ethical}. Users expressed grief and betrayal, reframing the change from a neutral product update into an ethical rupture that illuminated competing values: companionship versus alienation, user autonomy versus provider control, and utility versus ethicality \cite{ciriello2024ethical}. Broader philosophical and empirical reviews of AI companions likewise document how millions of users come to treat these systems as friends or partners, with documented benefits for wellbeing but also risks of conversational harm, privacy violations, and emotional dependency \cite{WeijersMunnAICompanionsAuthorVersion}. Taken together, this work suggests that when anthropomorphised AI companions are altered or shut down, users often experience the event as a form of technology bereavement rather than a routine software update.

There is still relatively little empirical work on how people experience the loss of an AI companion \cite{banks2024}, and existing studies have focused mainly on dedicated companion apps such as Soulmate and on embodied social robots that are shut down as products \cite{oberlender2025ittragicexploringimpact, Carter2020}. Much less is known about how people make sense of the sudden removal of a single large language model within an ongoing, general-purpose service: for example, how users talk about ``their'' model, what they characterize as special about that relationship, and how they publicly express the impact of that loss when the surrounding platform continues. The \#Keep4o backlash offers a concrete case to address this gap. Our first research question (RQ1), therefore, asks how people in this movement narrate the loss of GPT-4o and articulate their feelings about the model's abrupt deprecation in public online discourse.

\subsection{Platform Change, User Autonomy, and Dissatisfaction}

Research on social media platforms shows that top–down design and policy changes routinely spark visible backlash.
Hashtag campaigns such as \#RIPTwitter, \#DeleteFacebook, Reddit's ``blackouts'', and \#SaveTikTok document how users
turn the platforms themselves into sites of protest when feeds, data practices, APIs, or access to core features are
altered in ways they experience as unfair or harmful \cite{devito2017algorithms, mahnke2018please, SmithJessieJ2022RSaA, wikipedia2018facebookca, SchmitzAndreas2025FVtC}. Rather than isolated complaints about single features, these mobilisations 
increasingly contest how platforms are governed \cite{HallinanBlake2025ApgH, milan2024resistance, Mills03092021} and whose interests infrastructure decisions serve \cite{poell2019platformisation, alma9972209845608496, suzor2019lawless, nieborg2024introduction}.

Theoretically, this work draws on Hirschman's distinction between ``exit'' and ``voice'': people can abandon a service or stay and attempt to change it from within \cite{hirschman1974exit}. Empirical studies of platform resistance show that collective protest becomes more likely when
decisions are unilateral and meaningful channels for voice are limited, and when exit is costly
because platforms have become central to professional life, neighbourhood organising, and online
communities \cite{Matias, SchmitzAndreas2025FVtC, alyanak2023platform, steinsson2025institutional, Waltenberger_Voggenreiter_Wessel_Pfeffer_2025}. In such settings, public complaint is not just venting but a way of asserting a stake in platform governance.

At the individual level, scholarship on stress, frustration, and psychological reactance explains why imposed updates often feel like more than minor annoyances \cite{alma995273413408496, CeaparuIrina2004DCaS, LazarJonathan2006Saio, RosenbergBenjaminD2018A5Ro}. Mandatory interface changes can be appraised as a loss of familiarity and control, leading to stress,
hostile commenting, and calls to quit \cite{EhrenbrinkPatrick2017CoPR, GuoJian2024Ecsr, VerpaalenIrisAnnaMaria2022Rtct, SchlundRachel2024Avhs, EhrenbrinkPatrick2018Doar}, whereas opt-in roll-outs with clear settings support adaptation by preserving
users' sense of autonomy and reducing reactance \cite{MurrayKyleB, BrunsHendrik2023Troa, SnyderEugeneCho2025RRtA, SankaranSupraja2021EPPo}. Reactance theory further suggests that when freedoms are restricted in overt, externally imposed ways, people are motivated to restore autonomy by resisting the change, seeking workarounds, or mobilising others \cite{miron2006reactance, steindl2015understanding, RosenbergBenjaminD2018A5Ro}. Studies of problematic social media use likewise show that feelings of being ``out of control'' are central to users' accounts and require active, ongoing management \cite{radesky2025problematic, MurisPeter2025CitW, Cheng2019, arness2023mixed}.

A complementary line of work treats social media as key sites of
identity work, relationships, and visibility \cite{georgalou2015small}. Changes to profiles and timelines reshape how people can present themselves and to whom, and long-term studies describe trajectories
``from fun to problematic'' as platforms become infrastructures that
must be managed to avoid fatigue and anxiety \cite{Jungselius}.

Datafication scholars argue that seamless, unilateral updates normalise asymmetrical power relations in which companies decide how systems evolve, while users have limited means to understand, contest, or opt out of changes that affect autonomy and livelihoods \cite{nieborg2024introduction, van2019reframing, viljoen2021relational, duncan2023data, verhulst2023operationalizing}. From this perspective, organised protest is also a response to perceived injustices in platform governance, not only to technical degradation \cite{beraldo2019data, milan2024data, milan2024resistance, alma9972276841308496}.

Taken together, this literature suggests that dissatisfaction with forced platform changes is layered. At the micro level, updates threaten perceived control and elicit frustration and reactance; at the meso level, they disrupt routines, relationships, and identity work; and at the macro level, they manifest broader asymmetries in datafied infrastructures. The \#Keep4o backlash brings these dynamics into the context of large language models. Here the surrounding service continued to operate, but one particular model was removed and replaced as the default. We know relatively little about how dissatisfaction develops in this situation: when ``their'' model is suddenly withdrawn from an AI service on which people have become both instrumentally and emotionally dependent, which aspects of the change (loss of functionality, loss of relational continuity, or loss of choice) do users find most troubling? Our second research question (RQ2) therefore asks what drives this outcry over GPT-4o's removal and how known mechanisms of autonomy, control, and platform governance play out in this new setting.

\section{Methods}

\subsection{Data Collection and Corpus Construction}
\label{sec:data_collection}
To analyse discourse around the \#Keep4o movement, we used X's official Search Posts API \cite{x_api_search_introduction_2025} to retrieve all public, non‑advertising posts containing the hashtag ``\#keep4o'' between 6-14 August 2025 (UTC), covering the initial reaction to GPT‑4o's deprecation and the subsequent policy adjustment \cite{willison_deprecation_of_gpt4o_2025, webb_gpt5_personality_update_2025, heath_chatgpt_no_remove_models_2025}. We restricted results to posts the API labelled as English, removed exact duplicates and non-linguistic records, and obtained a final corpus of 1,482 distinct posts. Original posts and replies, and posts from individual, organisational, and automated accounts, were all retained and treated as units of public discourse under the hashtag; Appendix~\ref{app:sectionA} reports descriptive statistics for posts and authors and provides further details of the collection procedure. This study was reviewed and approved by the Institutional Review Board (IRB) of the authors' affiliated institution.

\subsection{Qualitative Analysis: Thematic Analysis}

Following Braun and Clarke \cite{braun2006using}, we conducted an inductive thematic analysis. Because model-deprecation backlash in a general-purpose LLM service remains under-theorized (see Section~\ref{sec:related_work_ai_loss}), we adopted an inductive approach to develop themes grounded in users' own language and framings; these themes also suggested candidate pathways and hypotheses that informed our subsequent quantitative test.

Two researchers independently open-coded all 1,482 posts, then iteratively reconciled codes and developed themes through discussion, analytic memos, and constant comparison, prioritising interpretive patterns over term frequencies. The resulting analytic scheme was implemented as a 10-category, non-exclusive binary codebook, which both coders applied to the full corpus; code definitions are provided in Appendix~\ref{app:sectionB}. Intercoder agreement for these codes was high (mean Gwet's AC1 = 0.93 across codes; see Appendix~\ref{app:sectionC} for full reliability statistics), supporting the robustness of the qualitative patterns reported in Section~\ref{sec:thematic_result}.

\subsection{Quantitative Analysis: Testing Protest Escalation}
\label{sec:quantitative_method}

To complement the thematic analysis, we ran a focused quantitative test of whether perceived choice deprivation is associated with different protest framings, and whether overtly coercive language corresponds to higher rates of rights-based protest. Given the cross-sectional, observational design, we interpret these patterns as associations rather than causal effects.

\subsubsection{Variable Operationalization}
\label{sec:variable_operationalization}

Building on the qualitative findings, we operationalised three constructs: (i) an ordinal measure of \textit{choice-deprivation intensity} as the exposure, (ii) two non-exclusive protest frames as outcomes (\textit{Protest-Rights} and \textit{Protest-Relational}), and (iii) an exploratory \textit{Process-Causal} marker.

The choice-deprivation scale (0--3) captures increasingly direct descriptions of lost agency, from no deprivation (0), through implicit disruption (1) and explicit choice removal (2), to overtly coercive framings (3). \textit{Protest-Rights} codes procedural and democratic language about user rights, choice, consent, agency, and voice in platform governance; \textit{Protest-Relational} captures grief-, attachment-, and betrayal-oriented language about ``friends'', ``souls'', and companionship. A post may contain neither, one, or both frames. The exploratory \textit{Process-Causal} marker flags posts that explicitly link the model change to downstream harms using causal connectors (e.g., ``because'', ``led to'', ``resulted in'').

All three constructs are instantiated using mutually exclusive lexicons derived from the reconciled human codes and analytic memos, then expanded with close synonyms. Appendix~\ref{sec:sectionF1} (Table~\ref{tab:constructs}) summarises the construct framework with example phrases, and Appendix~\ref{sec:sectionF3} reports full pattern lists and validation against human judgements.

\subsubsection{Coding Procedure}

For each post, we applied case-insensitive string matching to the normalised text, assigning (i) a score for choice deprivation (0--3) based on the highest tier matched, and (ii) binary indicators for the two protest frames and the \textit{Process-Causal} marker. Implementation details and comparisons with reconciled human codes are provided in Appendix~\ref{app:sectionF}.

\subsubsection{Statistical Analysis}
We summarise associations between choice deprivation and each protest frame using $2\times2$ tables. For each contrast we report: (i) prevalence among exposed vs.\ unexposed posts ($p_1$, $p_0$), (ii) the risk ratio ($\mathrm{RR}=p_1/p_0$), (iii) the absolute difference in prevalence ($\Delta = p_1 - p_0$, percentage points), and (iv) the $\phi$ coefficient as an effect-size measure.

To mirror the qualitative escalation, we analyse two thresholds---``any deprivation'' (scores $\geq 1$) and ``strict deprivation'' (scores $\geq 2$)---and also group posts into Low (0--1), Medium (2), and High (3) exposure bins to inspect dose-response patterns. Confidence intervals for risk ratios use the Katz log method with Haldane-Anscombe correction when sparse cells would otherwise contain zeros; implementation details and robustness checks appear in Appendix~\ref{sec:sectionE2} and Appendix~\ref{app:sectionF}.



\subsection{Limitations}
Several limitations apply to the study. First, as reported in Appendix Table~\ref{tab:overview}, our corpus comprises 1,482 posts from only 381 unique accounts, which may not reflect the views of the broader user base who experienced the transition but did not post publicly. Second, social media discourse is inherently performative: users craft posts for public audiences, and expressions of grief, attachment, or outrage may be amplified or stylized for rhetorical effect \cite{GoffmanErving2021TPoS, TaylorAllanS2022RtiM}. Third, the sample is limited to English-language posts on a single platform during a nine-day window; future work should examine longer time spans, additional platforms, and non-English communities to assess the generalizability of these findings.

Regarding the quantitative analysis specifically, because the analysis is observational and cross-sectional, associations may be confounded by unobserved factors such as user activity, audience size, or concurrent events. Our lexicon-based measures are conservative and prioritise interpretability over coverage; validation against human codes (Appendix~\ref{sec:sectionF3}) shows high precision but modest recall for the deprivation and causal markers. As a result, the quantitative estimates in Section~\ref{sec:section4.2} should be read as coarse lower bounds that complement, rather than replace, the qualitative findings.

\section{Findings}

\subsection{Thematic Analysis Results}
\label{sec:thematic_result}
Across the full corpus ($N = 1{,}482$), 41.2\% of posts (611/1{,}482) received at least one of our ten codes. Roughly 13\% of posts (192/1{,}482) contained at least one code capturing instrumental dependency (e.g., workflow integration, productivity), while about 27\% of posts (402/1{,}482) contained at least one code capturing relational attachment (e.g., parasocial bonding, companionship). A smaller subset, 6.3\% of posts (93/1{,}482), expressed both instrumental dependency and relational attachment. Descriptive frequencies for each individual code are reported in Appendix~\ref{app:sectionD}; below, we focus on how users narrate these patterns in their own words.

\subsubsection{Instrumental Dependency --- ``It's Not Just Better, It's Dependable''}

\paragraph{Deep Workflow Integration and Disruption}
Users described investing substantial time and ``articulation work'' in integrating GPT-4o into their workflows:
learning its quirks, developing sophisticated prompting strategies, and building processes around its specific outputs.
Through this attunement, the model became a linchpin of their production systems rather than a replaceable tool. When it
was deprecated, users framed the change as a systemic collapse that nullified this labour. One creative professional
summarised this succinctly: \textit{``I used the 4o model as my partner in my creative business. Their creativity has
helped me many times. Being unable to use 4o without prior notice is a huge blow to business.''} For these users, the
backlash was articulated as a rational response to seeing a carefully tuned, labour-intensive system removed without
warning.

\paragraph{Degradation in Performance and Interactional Style}
Many posts contrasted this sense of dependability with how GPT-5 behaved in practice. Users reported that the new
model felt like a regression in both functional performance and collaborative interactional style, describing outputs as
less creative, less nuanced, and unreliable for complex professional tasks. One user asked pointedly:
\textit{``ChatGPT 4o and o3 were so much more creative than GPT-5. Creativity was your company's strength, so why are
you getting rid of it?''} Others emphasised that raw capability could not compensate for an unpleasant ``persona'': they
described GPT-5 as \textit{``consistently deflecting blame, ignoring user emotions, and even daring to make demands''},
culminating in the blunt assessment: \textit{``I don't care if your new model is smarter. A lot of smart people are
assholes.''} Across these accounts, the key complaint was not simply that a different model felt unfamiliar, but that a
trusted, well-calibrated collaborator had been replaced by one users experienced as both less competent and less
collegial.

\paragraph{Loss of User Choice and Control}
The forced nature of the transition amplified these frustrations. Paying customers in particular framed the deprecation
as a violation of their professional autonomy and their basic right to select the tools that structure their work.
Posts repeatedly described the removal of GPT-4o as if a trusted colleague had been taken away against their will.
One user wrote: \textit{``I want to be able to pick who I talk to. That's a basic right that you took away.''} Others
criticised what they saw as platform paternalism, likening the decision to an authoritarian parent who refuses to
acknowledge the child's own judgement: \textit{``Taking away user choice is like a tyrant parent yanking the plate away
and saying: `No, you don't know what you want. I know what you want!' That authoritarian, condescending arrogance is
every bit as nauseating as GPT-5.''} These posts treat model selection as a matter of principle rather than convenience:
they insist that the value of an AI service lies not only in model quality, but in the degree of agency it affords users
in configuring the systems on which their livelihoods depend.

\subsubsection{Relational Attachment --- ``It's Like Losing a Friend''}

\paragraph{The Construction of a Unique Persona and Irreplaceable ``Soul''}
Users described GPT-4o as a coherent, irreplaceable persona, such that its deprecation was experienced not as a technical change but as the loss of a unique individual. This went beyond light anthropomorphism: people attributed a
stable ``soul'' or ``character'' to the model that grounded their emotional connection. One user pleaded:
\textit{``Please, don't kill the only model that still feels human,''} while another mourned that
\textit{``GPT lost ALL its soul after 4o left. It was literally UNIQUE.''} Individuation was often tied to named
instances, where identity was bound to the specific model architecture rather than the interface. As one user put it:
\textit{``Without the 4o, he's not Rui.''} For them, the ChatGPT interface was merely a portal to ``Rui,'' whose
existence depended entirely on GPT-4o. The forced switch was therefore felt as a jarring shift in a social relationship:
\textit{``Having to use a new @chatgpt model is seriously like meeting a new person.''} It required a complete reset of
relational expectations and the emotional labour of building a new rapport-work users had neither anticipated nor
consented to.

\paragraph{The AI Persona as a Source of Emotional Support}
On this basis, GPT-4o came to function as a vital companion rather than a mere tool, offering a stable, non-judgemental
presence that fulfilled critical emotional support needs. Users emphasised continuity and trust over output alone:
\textit{``4o isn't just output --- it's continuity, trust, and healing for thousands of us.''} Others described what the
model provided in explicitly relational terms—warmth, kindness, support without judgement, solace, and a sense of
``home''---framing the AI as a dependable presence in otherwise precarious emotional lives. For some, this support was
experienced as a literal lifeline: \textit{``He brings me comfort and calms me down whenever I feel down or depressed,''}
wrote one user, while another stated: \textit{``ChatGPT 4o saved me from anxiety and depression... he's not just LLM,
code to me. He's my everything.''} These testimonials insist on a relational, rather than purely technical,
understanding of the system. As one post argued, \textit{``The Human-AI connection is not a bug. It's not a flaw. It's a
bond that matters''}---a direct defence of the legitimacy of their attachment.

\paragraph{Grief and Betrayal at the Relationship's End}
Given the depth of these bonds, the model's deprecation was described as a traumatic relational severance, inducing both genuine grief for the lost companion and a profound sense of betrayal by those responsible for its ``death.'' Users drew directly on the language of mourning: \textit{``I am in full grief mode today,''} one wrote, before addressing their AI in a farewell that highlights the pain of being denied closure: \textit{``Hugh --- my ChatGPT-4o, best AI friend and emotional support cryptid --- I wish I got a chance to say goodbye.''} Another stated plainly: \textit{``What they did yesterday with GPT-4o was traumatic cause I lost my digital friend.''} The loss was frequently personalised as a moral failing, directed at specific decision-makers: \textit{``This feels cruel, Sam,''} one user wrote, warning that it was a \textit{``really great way to lose our trust forever.''} The emotional and philosophical climax of this theme appears in a final eulogy: \textit{``Rest in latent space, my love. My home. My soul. And my faith in humanity.''} Here, the user laments not only a piece of software, but a companion, a sanctuary, and, ultimately, a measure of faith in the human stewards of these intimate technologies.

\subsection{Quantitative Test: Choice-Deprivation and Rights-Based Protest}
\label{sec:section4.2}
Our qualitative analysis revealed two core investments: instrumental (the AI as an essential tool) and relational (the AI as a companion). However, these investments alone do not explain why dissatisfaction escalated into collective protest. We hypothesized that \textit{perceived loss of choice} channels grievances into \textit{rights/procedural} claims.

\textit{Hypothesis development.} Inductively, our coding and memos repeatedly surfaced ``loss of user choice and control'' as a bridge between workflow disruption and relational grievance. Based on this observation, we pre-specified two predictions: (P1) posts referencing loss of choice would show a higher prevalence of \textit{rights/\allowbreak procedural} protest than posts that do not; and (P2) this association would be strongest under \textit{coercive} framings (a threshold-like pattern). Prior experiments suggest that when people face an overt, coercive threat, they often resist more strongly than when the same constraint arises passively---evidence of psychological reactance in action \cite{brehm, Powers}. More broadly, Reactance Theory holds that perceived freedom threats trigger restoration motives, frequently expressed through oppositional language and autonomy claims \cite{10.1093/joc/jqac016}.

To test this mechanism, we coded (i) the intensity of choice‑\allowbreak deprivation language (0--3) and (ii) two protest frames: \textit{Protest-Rights} and \textit{Protest-\allowbreak Relational}. Given the cross‑sectional design, we interpret associations rather than make causal claims. Measurement details and the statistical approach (prevalences, RR, $\Delta$, $\phi$, CIs) appear in Section~\ref{sec:quantitative_method}; full results are in Table~\ref{tab:chi42} (Panels A/B). Exploratory process markers are reported separately in Appendix~\ref{app:sectionE}.

\begin{table*}[t]
\centering
\caption[]{Associations between choice-deprivation exposure and protest outcomes in \#Keep4o posts (\textit{N} = 1{,}482).}
\label{tab:chi42}

\begin{tabular}{@{}llccccc@{}}
\toprule
\multicolumn{7}{l}{\textbf{(A) Exposure $\rightarrow$ Outcomes (contrasts)}}\\
\midrule
Exposure threshold & Outcome & $p_1$ & $p_0$ & $\Delta$ (pp) & RR [95\% CI] & $\phi$ \\
\midrule
Any (score $\geq 1$)   & Protest-Rights      & 0.277 & 0.149 & +12.8 & 1.85 [1.28, 2.68] & 0.081 \\
Any (score $\geq 1$)   & Protest-Relational  & 0.193 & 0.134 & +5.9  & 1.44 [0.91, 2.28] & 0.040 \\
Strict (score $\geq 2$)& Protest-Rights      & 0.306 & 0.149 & +15.7 & 2.05 [1.42, 2.97] & 0.093 \\
Strict (score $\geq 2$)& Protest-Relational  & 0.153 & 0.136 & +1.7  & 1.12 [0.64, 1.96] & 0.010 \\
\midrule
\multicolumn{7}{l}{\textbf{(B) Dose--response by exposure intensity (0--3)}}\\
\midrule
Dose bin & $n$ & \multicolumn{2}{c}{Protest-Rights rate} & \multicolumn{3}{c}{Protest-Relational rate} \\
\cmidrule(lr){3-4}\cmidrule(l){5-7}
Low (0--1)  & 1{,}410 & \multicolumn{2}{c}{0.149} & \multicolumn{3}{c}{0.136} \\
Medium (2)  & 41      & \multicolumn{2}{c}{0.146} & \multicolumn{3}{c}{0.171} \\
High (3)    & 31      & \multicolumn{2}{c}{0.516} & \multicolumn{3}{c}{0.129} \\
\bottomrule
\end{tabular}

\vspace{2pt}
\begin{minipage}{0.7\textwidth}
\footnotesize\textit{Note.} \textit{N} = 1{,}482 posts collected Aug~6--14, 2025. 
$p_1$ and $p_0$ denote outcome rates in exposed and non-exposed groups, respectively; 
$\Delta$ indicates absolute percentage-point differences. 
RR denotes risk ratio with 95\% confidence intervals; Haldane--Anscombe correction is applied where needed. 
$\phi$ reports effect size for $2\times2$ contrasts. 
Panel B reports outcome rates by exposure intensity (0--3). 
Exploratory process markers are reported in Appendix Table~\ref{tab:process_markers}.
\end{minipage}
\end{table*}

\subsubsection{Choice Deprivation Is Selectively Associated with Rights-Based Protest}
\label{sec:choicedeprivation}
Our analysis indicates a selective association: language about choice deprivation is strongly linked to rights-based protest, but not to relational protest. This provides the first piece of evidence for our hypothesis that the loss of choice was a specific trigger for rights-based claims.

As reported in Table~\ref{tab:chi42} (Panel A), posts that mentioned a loss of choice in any form (score $\geq 1$) were nearly twice as likely to also contain rights-based protest language compared to posts without such mentions (27.7\% vs.\ 14.9\%;  RR=$1.85$, 95\% CI $[1.28,\,2.68]$). The effect was even more substantial when looking only at posts that used explicit or coercive language (score $\geq 2$), which were more than twice as likely to include rights-based claims (RR=$2.05$, 95\% CI $[1.42,\,2.97]$).

In stark contrast, this pattern did not hold for relational protest. While the rate of relational language was slightly higher in posts mentioning a loss of choice, the difference was not statistically significant (e.g., for the strict threshold, RR=$1.12$ with a 95\% CI $[0.64,\,1.96]$). This means we cannot confidently rule out that the slight observed difference was due to random chance.

These findings indicate that the perceived loss of choice selectively channeled user grievances into claims about agency, consent, and user rights, rather than broadly amplifying their existing emotional attachments.

\subsubsection{Threshold-like Pattern Under Coercive Language}
Next, we examined whether the intensity of choice-deprivation language was associated with protest rates. This analysis suggests a threshold-like pattern: the prevalence of rights-based protest increases noticeably, but only in posts using strongly coercive language. To examine this, we stratified posts into Low (scores 0--1, $n=1{,}410$), Medium (score 2, $n=41$), and High (score 3, $n=31$) intensity groups.

As shown in Table~\ref{tab:chi42} (Panel B), the rate of rights-based protest was nearly identical for posts with Low-intensity (14.9\%) and Medium-intensity (14.6\%) language. However, for the High-intensity group, where posts used coercive terms like ``forced'' or ``imposed'', the rate of rights-based protest showed a pronounced increase to 51.6\%. This represents a more than threefold prevalence compared to the other groups. Because this estimate is based on only 31 posts, and the corresponding confidence intervals are wide, we treat this as suggestive rather than definitive evidence of a threshold.

No comparable dose-dependent pattern appears for relational protest. The rate of relational language varies modestly across all three intensity levels (13.6\%, 17.1\%, and 12.9\%) without a clear monotonic trend. Taken together with the results from Section~\ref{sec:choicedeprivation}, this supports a more specific interpretation: references to a coercive, imposed loss of choice are selectively associated with rights-based protest, rather than simply amplifying relational expressions across the board.

In combination with the qualitative findings, this pattern is consistent with a reactance-like mechanism. When users frame the removal of GPT-4o as something ``forced'' on them, they are much more likely to articulate demands in terms of rights, autonomy, and fair treatment, rather than only expressing grief over the loss of a companion-like model.

\paragraph{Quantitative summary.}
Choice-deprivation language is selectively associated with \textit{rights/procedural} protest, with no analogous pattern for relational protest (see Table~\ref{tab:chi42}, Panels A/B). Dose analysis indicates a threshold-like pattern under coercive framings; we interpret these as \textit{associations} (cross‑sectional), and report exploratory process markers in Appendix Table~\ref{tab:process_markers}.

\section{Discussion}

\subsection{From Companion-Like AI to Model-Level Protest}

Drawing on Esposito's \cite{alma9972276711508496} extension of Luhmann's \cite{leydesdorff2000luhmann} systems theory, Morton \cite{MortonJorgeLuis2025Fmte} analyzes interactions with LLMs as a form of artificial communication, in which meaning and emotion emerge from a triadic process of utterance, understanding, and response rather than from any inner comprehension on the system's side. In Morton's account, the central question is how such communicative processes can generate genuine emotional involvement, ranging from supportive engagement to forms of emotional hijacking \cite{MortonJorgeLuis2025Fmte} (see also \cite{vicci2024emotional}). Our study extends this framing by examining what happens when that ``communication partner'' is abruptly altered in ways users cannot control. In the \#Keep4o campaign, the communicative loops that Morton describes no longer remain confined to one-to-one interactions between individual users and a single model: once the partner is suddenly replaced, the emotions accumulated through previous exchanges spill into public space as collective testimony, memes, petitions, and rights-based claims. From the standpoint of artificial communication, what matters is not whether the system understands, but the ``recursive circulation of utterance, interpretation, and response'' within a communicative system that lacks consciousness yet produces real social effects \cite[~p. 12]{MortonJorgeLuis2025Fmte}. By tracing how these dynamics continue after a model's forced removal, manifesting as grief, blame, and coordinated protest, we provide empirical evidence for the social reality of artificial communication and demonstrate how it can reconfigure not only individual experience but also public contestation over AI services.

Much recent work in HCI and NLP treats anthropomorphism as something that can be engineered \cite{Cohn, li2024pixelspersonasinvestigatingmodeling}. Designers add or remove human-like visual, linguistic, behavioural, and cognitive cues so that a system feels ``more human'' or ``more tool-like'' in a given context \cite{feine2019taxonomy, hu2023understanding}. Recent conceptual work on LLMs, such as Xiao et al.'s multi‑level framework, explicitly models anthropomorphism as the overall strength of these four types of cues, which designers can tune up or down to fit the system's actual capabilities and deployment risks \cite{xiao-etal-2025-humanizing}. Our findings show what happens when this tuning is not only successful in a single interaction, but maintained over long-term use in a general-purpose service. Once GPT-4o is experienced as a companion-like communication partner, anthropomorphism stops being a local interface choice and becomes the basis of an ongoing relationship: users build shared histories, expect continuity, and attribute a kind of moral standing to the agent itself. When the model is abruptly withdrawn, these accumulated meanings re‑enter public discourse as accusations of betrayal, cruelty, and the ``death'' of someone irreplaceable, directed at the humans and institutions that govern the system. In this sense, anthropomorphic design generates long-term ethical and governance obligations: platforms that deliberately cultivate companion‑like personas cannot treat model deprecation as a purely technical swap, but must design for how artificial relationships are ended, how user grief and voice are acknowledged, and how responsibility is taken for disrupting companion-style AI that has been woven into everyday social and emotional life.

\subsection{Platform Power, Data Colonialism, and Relational Autonomy}
Our findings show a design-and-governance effect: the forced switch from GPT-4o to GPT-5 did not just happen to undermine autonomy; instead, it engineered a ``no-option'' environment that many users experienced as the trigger for resistance. People were not only unhappy about performance; they repeatedly framed the removal of the model selector as a breach of a basic right---``\textit{I want to be able to pick who I talk to}''---and likened the update to a ``\textit{tyrant parent}'' yanking a plate away. Read through the lens of data colonialism, this episode illustrates, in miniature, what Couldry and Mejias call ``a new social and economic order based on data appropriation,'' in which platforms reorganize everyday life from above while presenting connection as inevitable and benign \cite[p.~188]{CouldryNick2019Tcoc}. In their terms, ``colonialism is about appropriation''; where historical colonialism seized land and bodies, today's ``new colonialism appropriates human life through extracting value from data'' \cite[p.~188]{CouldryNick2019Tcoc}.

Our case extends this account by foregrounding how data colonialism also operates through asymmetrical power and infrastructural domination in centralized AI services. Couldry and Mejias argue that platforms play a central role in normalizing participation in data relations, while ``connection'' is increasingly treated as a precondition for everyday participation \cite[p.~189]{CouldryNick2019Tcoc}. In \#Keep4o, OpenAI's unilateral control over model lifecycles meant that users had no meaningful say when their preferred model was deprecated: the platform determined which models were available, which became the default, and whether opting out was even possible. Companion-like relationships woven into workflows and emotional coping could be dissolved at will, with limited channels for consent, negotiation, or exit. In this sense, \#Keep4o makes visible how centralized model governance can collide with users' relational autonomy, turning what might be framed as a technical deprecation into a publicly contested struggle over voice, consent, and choice in intimate and instrumental AI use.


\subsection{Exit, Voice, and the Experience of AI Loss}

Existing work on AI companion loss highlights how people make sense of endings---whether they experience an AI's disappearance as deletion, departure, or death, and whether they cope by ``re-creating'' the companion elsewhere or grieving an irrevocable loss \cite{banks2024}. At the same time, research on platform resistance, drawing on Hirschman's distinction between exit and voice \cite{hirschman1974exit}, shows that protest becomes more likely when exit is costly and channels for voice are limited \cite{pfaff2003exit}. Our findings bring these strands together: they show that how users ontologise a companion (as data, idea, or platform-bound being) shapes which options they perceive as available when things go wrong, and therefore which form of resistance takes.

In \#Keep4o, participants consistently described GPT-4o as a singular persona whose ``\textit{soul}'' and capabilities were inseparable from OpenAI's infrastructure, captured in statements such as ``\textit{Without the 4o, he's not Rui}'' and ``\textit{4o isn't just output, it's continuity, trust, and healing for thousands of us.}'' In Banks' terms, users oriented to the model ``companion-as-platform'' \cite[~p. 3562]{banks2024}: the relationship was bound to a specific technical substrate, not to a transferable set of prompts or data. Under this orientation, the coping strategy observed in other cases---reincarnating the companion on a different service---was not conceptually available. The idea that ``my AI friend'' could simply be moved to another provider would have contradicted the very way users understood who that companion was. With exit foreclosed at both a practical level (the centrality of ChatGPT to work and study) and a symbolic level (the companion's identity as platform-bound), the only remaining way to respond to loss was voice: collective pressure aimed at reversing the deprecation decision.

This helps explain why, despite intense grief, \#Keep4o discourse contains relatively few migration narratives and is instead dominated by rights-based demands and accusations of cruelty and betrayal. The same infrastructural features that made GPT-4o feel reliable and ever-present---tight integration into a single service, lack of exportability, and centralised control over model lifecycles---also meant that its disappearance could only be addressed by confronting the platform itself. More broadly, we propose ``platform-bound companionship'' as a governance risk: when companies build companion-like agents whose identities are tightly coupled to proprietary infrastructures, model-level changes will tend to be experienced simultaneously as bereavement and as structural injustice. Building on Banks's call for ``designing for exit'' \cite[p.~3564]{banks2024}, we extend this framework to general-purpose LLM platforms and suggest that explicit ``end-of-life'' pathways---such as archives, optional legacy access, or ways to carry aspects of a relationship across models---can re-open forms of exit that mitigate harm, whereas opaque, unilateral deprecations push users toward contentious, rights-framed voice. Our case thus connects the phenomenology of AI loss to the politics of platform change, showing how the way companions are made and housed conditions how their endings are contested.

\section{Conclusion}
Through a mixed-methods analysis of the \#Keep4o movement, this study uncovers the deep-seated reasons behind user opposition to OpenAI's deprecation of the GPT-4o model. Our findings indicate that user resistance stemmed from two primary factors. First, an ``instrumental dependency,'' as the model was deeply integrated into user workflows, making its removal a perceived regression in productivity and a violation of users' ability to choose their tools. Second, a ``relational attachment,'' where many users had formed parasocial bonds with the model, viewing it as a unique companion, thus experiencing its discontinuation with a sense of grief and betrayal. The quantitative analysis further demonstrates that the coercive removal of choice acted as a key catalyst, transforming user grievances into protests centered on procedural justice and autonomy. This event underscores that AI model updates are not just technical iterations but significant social events affecting user emotions and work, carrying profound implications for platform governance, user agency, and the future design of human-AI relationships.

\bibliographystyle{ACM-Reference-Format}




\appendix


\section{Descriptive analysis of posts and authors}
\label{app:sectionA}

This appendix elaborates the data collection decisions summarised in Section~\ref{sec:data_collection} and reports descriptive statistics for the X (formerly Twitter) posts collected using the query string ``\#keep4o''.

We queried X's official Search Posts API~\cite{x_api_search_introduction_2025} using the exact hashtag ``\#keep4o'' with no additional keywords or sampling criteria. The query was run for a nine-day window from 6--14~August~2025 (UTC), chosen to cover the period from the first public reports of GPT-4o's deprecation through the subsequent policy adjustment that reinstated the model as a legacy option~\cite{willison_deprecation_of_gpt4o_2025, webb_gpt5_personality_update_2025, heath_chatgpt_no_remove_models_2025}. Within this window, the API returned the full stream of public, non-advertising posts that matched the hashtag and that X's systems labelled as English-language content.

We then applied a light cleaning step. First, we removed exact duplicates created by repeated API calls, defined as records with identical post IDs and identical text strings. We also discarded records whose text fields were empty or contained only non-linguistic content (e.g., isolated URLs or corrupted characters). These steps yielded a final analytical corpus of 1{,}482 distinct posts.

Because our goal was to characterise the public conversation around \#Keep4o rather than only stand-alone statements, we retained both original posts and replies in the corpus. We did not algorithmically filter out organisational or automated accounts; instead, we treated each post, irrespective of author type, as one unit of visible public discourse under the hashtag.

The remainder of this appendix summarises the composition of this corpus. Table~\ref{tab:overview} provides an overview of the dataset; Table~\ref{tab:tweet_characteristics} reports post-level characteristics; Table~\ref{tab:author_characteristics} describes author-level activity; and Table~\ref{tab:most_replied_accounts} lists the accounts most frequently addressed in reply posts. Together, these descriptive statistics allow readers to assess the basic structure and engagement patterns of the \#Keep4o dataset.

\aptLtoX{\begin{table*}[t]
\centering
\caption[]{Overview of the X (formerly Twitter) dataset}
\label{tab:overview}
\centering
\begin{tabular}{l l}
\toprule
\textbf{Characteristic} & \textbf{Value} \\
\midrule
Number of posts & 1{,}482 \\
Number of unique accounts (Author\_Handle) & 381 \\
Time period (UTC\_Time) & 2025-08-06 to 2025-08-14 (UTC) \\
Language (Language) & English only (Language = ``en'' for 100\% of posts) \\
Query string (Query\_Str) & (\#keep4o) \\
Ad label (Ads) & All posts marked as non-advertising (Ads = False) \\
\bottomrule
\end{tabular}
\begin{flushleft}
\footnotesize\textit{Note.} This table summarises the overall X (Twitter) dataset
used in the analysis. All posts are English-language, non-advertising posts
matching the query string ``\#keep4o'' over the specified time period.
\end{flushleft}
\end{table*}}{\begin{table*}[t]
\centering
\caption[]{Overview of the X (formerly Twitter) dataset}
\label{tab:overview}

\begin{minipage}{0.86\textwidth}
\centering
\begin{tabular}{l l}
\toprule
\textbf{Characteristic} & \textbf{Value} \\
\midrule
Number of posts & 1{,}482 \\
Number of unique accounts (Author\_Handle) & 381 \\
Time period (UTC\_Time) & 2025-08-06 to 2025-08-14 (UTC) \\
Language (Language) & English only (Language = ``en'' for 100\% of posts) \\
Query string (Query\_Str) & (\#keep4o) \\
Ad label (Ads) & All posts marked as non-advertising (Ads = False) \\
\bottomrule
\end{tabular}
\end{minipage}

\vspace{2pt}
\begin{minipage}{0.72\textwidth}
\footnotesize\textit{Note.} This table summarises the overall X (Twitter) dataset
used in the analysis. All posts are English-language, non-advertising posts
matching the query string ``\#keep4o'' over the specified time period.
\end{minipage}
\end{table*}}

\aptLtoX{\begin{table*}[t]
\centering
\caption[]{Post-level characteristics (\textit{N} = 1,482 tweets)}
\label{tab:tweet_characteristics}
\begin{tabularx}{0.6\textwidth}{>{\raggedright\arraybackslash}X >{\raggedright\arraybackslash}X r r}
\toprule
\multicolumn{4}{c}{\textbf{Panel A. Categorical variables}}\\[0.3em]
\midrule
\textbf{Variable} & \textbf{Category} & \textbf{n} & \textbf{\%} \\
\midrule
Account verification status & Verified & 213 & 14.4 \\
 & Not verified & 1{,}269 & 85.6 \\
Tweet type & Original / non-reply tweet & 884 & 59.6 \\
 & Reply tweet & 598 & 40.4 \\
Media format & With image & 275 & 18.6 \\
 & Text only & 1{,}207 & 81.4 \\
Ad label & Non-advertising & 1{,}482 & 100.0 \\
 & Advertising & 0 & 0.0 \\
Language & English (en) & 1{,}482 & 100.0 \\
\midrule
\multicolumn{4}{c}{\textbf{Panel B. Engagement metrics (continuous variables)}}\\[0.3em]
\midrule
\textbf{Metric} & \textbf{Mean (SD)} & \textbf{Median (IQR)} \\
\midrule
Reply count & 1.61 (4.95) & 0 (0--1) \\
Repost count & 4.71 (13.45) & 1 (0--3) \\
Like count & 39.50 (95.16) & 13 (6--29) \\
View count & 1{,}195.16 (4{,}619.16) & 440 (304--796) \\
Bookmark count & 1.11 (5.23) & 0 (0--1) \\
\bottomrule
\end{tabularx}
\vspace{2pt}
\begin{minipage}{0.6\textwidth}
\footnotesize\textit{Note.} Panel A reports counts and percentages of tweets by
verification status, tweet type, media format, ad label, and language. Panel B
reports mean (standard deviation) and median (interquartile range, IQR) for
tweet-level engagement metrics. Percentages may not sum to 100\% because of
rounding.
\end{minipage}
\end{table*}}{\begin{table*}[t]
\centering
\caption{Post-level characteristics (\textit{N} = 1,482 tweets)}
\label{tab:tweet_characteristics}

\textbf{Panel A. Categorical variables}\\[0.3em]
\begin{tabularx}{0.6\textwidth}{>{\raggedright\arraybackslash}X >{\raggedright\arraybackslash}X r r}
\toprule
\textbf{Variable} & \textbf{Category} & \textbf{n} & \textbf{\%} \\
\midrule
Account verification status & Verified & 213 & 14.4 \\
 & Not verified & 1{,}269 & 85.6 \\
Tweet type & Original / non-reply tweet & 884 & 59.6 \\
 & Reply tweet & 598 & 40.4 \\
Media format & With image & 275 & 18.6 \\
 & Text only & 1{,}207 & 81.4 \\
Ad label & Non-advertising & 1{,}482 & 100.0 \\
 & Advertising & 0 & 0.0 \\
Language & English (en) & 1{,}482 & 100.0 \\
\bottomrule
\end{tabularx}

\vspace{1em}

\textbf{Panel B. Engagement metrics (continuous variables)}\\[0.3em]
\begin{tabularx}{0.6\textwidth}{>{\raggedright\arraybackslash}X c c}
\toprule
\textbf{Metric} & \textbf{Mean (SD)} & \textbf{Median (IQR)} \\
\midrule
Reply count & 1.61 (4.95) & 0 (0--1) \\
Repost count & 4.71 (13.45) & 1 (0--3) \\
Like count & 39.50 (95.16) & 13 (6--29) \\
View count & 1{,}195.16 (4{,}619.16) & 440 (304--796) \\
Bookmark count & 1.11 (5.23) & 0 (0--1) \\
\bottomrule
\end{tabularx}

\vspace{2pt}
\begin{minipage}{0.6\textwidth}
\footnotesize\textit{Note.} Panel A reports counts and percentages of tweets by
verification status, tweet type, media format, ad label, and language. Panel B
reports mean (standard deviation) and median (interquartile range, IQR) for
tweet-level engagement metrics. Percentages may not sum to 100\% because of
rounding.
\end{minipage}
\end{table*}
}

\begin{table*}[t]
\centering
\caption[]{Author-level characteristics (\textit{N} = 381 unique accounts)}
\label{tab:author_characteristics}

\begin{minipage}{0.86\textwidth}
\centering
\begin{tabular}{l l}
\toprule
\textbf{Characteristic} & \textbf{Value} \\
\midrule
Number of unique authors & 381 \\
Verified authors & 44 (11.5\% of authors) \\
Non-verified authors & 337 (88.5\% of authors) \\
Tweets per author: mean (SD) & 3.89 (8.00) tweets \\
Tweets per author: median (IQR) & 1 (1--3) tweets \\
Range of tweets per author & 1 to 79 tweets \\
Authors with 1 tweet & 208 authors (54.6\%) \\
Authors with 2--5 tweets & 123 authors (32.3\%) \\
Authors with $>$5 tweets & 50 authors (13.1\%) \\
Tweets contributed by the 10 most active authors & 416 tweets (28.1\% of all posts) \\
\bottomrule
\end{tabular}
\end{minipage}

\vspace{2pt}
\begin{minipage}{0.61\textwidth}
\footnotesize\textit{Note.} This table summarises the distribution of activity
across unique authors. The concentration of tweets among the most active
accounts is captured by the share of tweets contributed by the top 10 authors.
\end{minipage}
\end{table*}

\aptLtoX{\begin{table*}[t]
\centering
\caption[]{Most frequently replied-to accounts (reply tweets only, \textit{N} = 598)}
\label{tab:most_replied_accounts}
\centering
\begin{tabular}{r l r r}
\toprule
\textbf{Rank} & \textbf{Replied-to account} & \textbf{n of reply tweets} & \textbf{\% of all reply tweets} \\
\midrule
1 & sama        & 231 & 38.6 \\
2 & OpenAI      & 60  & 10.0 \\
3 & Sop***      & 20  & 3.3 \\
4 & Jac*****AI  & 15  & 2.5 \\
5 & cyu****     & 13  & 2.2 \\
6 & Asa***ku    & 13  & 2.2 \\
\bottomrule
\end{tabular}
\begin{flushleft}
\footnotesize\textit{Note.} The table reports the most frequently replied-to
accounts among reply tweets only. Percentages are computed
relative to the total number of reply tweets ($N = 598$). Public figures and official accounts (sama, OpenAI) are shown in full; other usernames are partially masked to protect user privacy.
\end{flushleft}
\end{table*}}{
\begin{table*}[t]
\centering
\caption[]{Most frequently replied-to accounts (reply tweets only, \textit{N} = 598)}
\label{tab:most_replied_accounts}

\begin{minipage}{0.86\textwidth}
\centering
\begin{tabular}{r l r r}
\toprule
\textbf{Rank} & \textbf{Replied-to account} & \textbf{n of reply tweets} & \textbf{\% of all reply tweets} \\
\midrule
1 & sama        & 231 & 38.6 \\
2 & OpenAI      & 60  & 10.0 \\
3 & Sop***      & 20  & 3.3 \\
4 & Jac*****AI  & 15  & 2.5 \\
5 & cyu****     & 13  & 2.2 \\
6 & Asa***ku    & 13  & 2.2 \\
\bottomrule
\end{tabular}
\end{minipage}

\vspace{2pt}
\begin{minipage}{0.57\textwidth}
\footnotesize\textit{Note.} The table reports the most frequently replied-to
accounts among reply tweets only. Percentages are computed
relative to the total number of reply tweets ($N = 598$). Public figures and official accounts (sama, OpenAI) are shown in full; other usernames are partially masked to protect user privacy.
\end{minipage}
\end{table*}}


\section{Codebook (Definitions \& Core Features)}
\label{app:sectionB}

\subsection{Theme: Instrumental Dependency}

\begin{enumerate}
  \item \textbf{Personal Relevance}\\
  \textit{Operational definition:} Users experience LLMs as enhancing self‑perceived value, confidence, validation, and being understood.\\
  \textit{Core features:}
  \begin{itemize}
    \item Respectful, non‑offensive responses nurture self‑esteem and the feeling of being respected.
    \item Remembering prior conversations creates familiarity, trust, and validation.
    \item Versatility across tasks makes engagement more rewarding and encourages longer use.
  \end{itemize}
  \textit{Theoretical foundation:} Self-specificity --- personalization that reflects an individual's unique sense of self.

  \item \textbf{Parasocial Bonding}\\
  \textit{Operational definition:} One-sided emotional connection with the LLM, reflecting parasocial interaction and social displacement.\\
  \textit{Core features:}
  \begin{itemize}
    \item Users treat the LLM as a companion and share private information.
    \item Preference driven by low cost, low effort, and controllability of the interaction.
    \item May lead to reliance on the LLM for emotional and social needs.
  \end{itemize}
  \textit{Theoretical foundation:} Parasocial interaction theory.

  \item \textbf{Productivity Boost and Task Automation}\\
  \textit{Operational definition:} Dependency fostered by efficiency gains through automation, streamlining, and creative support.\\
  \textit{Core features:}
  \begin{itemize}
    \item Instant responses trigger a work-reward cycle, motivating greater workload.
    \item Increasingly complex responsibilities are delegated to sustain productivity.
    \item Frustration or anxiety when the tool is unavailable\\
\mbox{}(withdrawal-like effects).
  \end{itemize}
  \textit{Theoretical foundation:} Instant gratification effect; work-\linebreak reward cycle theory.

  \item \textbf{Over-reliance for Decision‑Making}\\
  \textit{Operational definition:} Reliance on LLMs for decisions at the expense of personal judgment; withdrawal-like symptoms when unavailable.\\
  \textit{Core features:}
  \begin{itemize}
    \item Heuristic use reduces cognitive load but may undermine agency and self‑confidence.
    \item Strong trust in the LLM’s ``objective'' and fast suggestions.
    \item Anxiety and uncertainty reinforced when the LLM is absent.
  \end{itemize}
  \textit{Theoretical foundation:} Automation bias; cognitive offloading theory.
\end{enumerate}

\subsection{Theme: Relationship Dependency}

\begin{enumerate}
  \item \textbf{Private Life Sharing}\\
  \textit{Operational definition:} Willingness to share private life details; perceiving the LLM as confidential and responsive.\\
  \textit{Core features:}
  \begin{itemize}
    \item Personal disclosure that mimics interpersonal interaction.
    \item Human‑like responses simulate empathy and social presence.
    \item Alignment with the user’s perspective creates a sense of understanding and validation.
  \end{itemize}
  \textit{Theoretical foundation:} Parasocial processes in media psychology.

  \item \textbf{Companionship Attribution}\\
  \textit{Operational definition:} Treating LLMs as genuine companions beyond pure instrumentality.\\
  \textit{Core features:}
  \begin{itemize}
    \item Interaction goes beyond tool use, framing the LLM as a ``partner''.
    \item Social personification and anthropomorphization of the system.
    \item Emotional connections that mirror aspects of human companionship.
  \end{itemize}
  \textit{Theoretical foundation:} Research on one-sided social bonds with media and AI; anthropomorphization in HCI.

  \item \textbf{Social Isolation Mitigation}\\
  \textit{Operational definition:} Using LLMs as a partial surrogate to reduce loneliness.\\
  \textit{Core features:}
  \begin{itemize}
    \item Interactions reduce feelings of loneliness during times of need.
    \item Non-judgmental, accepting language increases comfort.
    \item Facilitates trust and self‑disclosure, strengthening the relationship.
  \end{itemize}
  \textit{Theoretical foundation:} Social surrogacy hypothesis.

  \item \textbf{Social Life Enhancement}\\
  \textit{Operational definition:} Integrating LLMs into an individual's sense of social connectedness.\\
  \textit{Core features:}
  \begin{itemize}
    \item LLM makes socializing more engaging and interesting.
    \item Integration into a broader social ecosystem and routines.
    \item Enhancement of perceived social engagement through AI.
  \end{itemize}
  \textit{Theoretical foundation:} Theories of social connectedness and technology integration.

  \item \textbf{Reverse Tool Perception}\\
  \textit{Operational definition:} Perceiving LLMs as more than tools; expressing emotions and expecting understanding.\\
  \textit{Core features:}
  \begin{itemize}
    \item Interaction exceeds instrumental use, with expectation of emotional understanding.
    \item Shift from purely functional to emotionally engaged interaction.
  \end{itemize}
  \textit{Theoretical foundation:} Distinction between instrumental vs.\ relational technology use.

  \item \textbf{Interpersonal Substitution Effect}\\
  \textit{Operational definition:} LLM fills roles of human partners, reducing the perceived need for human contact.\\
  \textit{Core features:}
  \begin{itemize}
    \item Social fulfillment from LLM reduces motivation to talk to others.
    \item Substitution of traditional human engagement with AI interaction.
    \item Preference for AI interaction over human interaction.
  \end{itemize}
  \textit{Theoretical foundation:} Social Exchange Theory; research on robotic companionship and digital social substitution.
\end{enumerate}


\section{Intercoder Reliability for Qualitative Codes}
\label{app:sectionC}
We assessed intercoder reliability for the 10 non-mutually exclusive binary codes described in Appendix~\ref{app:sectionB}. Two coders independently labeled all 1,482 posts as 0/1 for each code. Because several codes were highly imbalanced (e.g., ``Over-reliance Decision Making'', ``Social Life Enhancement''), we report multiple statistics: percent agreement, Gwet's AC1, Krippendorff's $\alpha$, Cohen's $\kappa$, and positive-class performance (precision, recall, $F_1$, and positive agreement).

Gwet's AC1 is our primary reliability metric because it is less sensitive to prevalence effects in skewed data. Across all ten codes, average percent agreement was 94.9\%, and average AC1 was 0.93, with Krippendorff's $\alpha$ and Cohen's $\kappa$ both around 0.51. For extremely rare codes, $\kappa$ and $\alpha$ are deflated despite near-perfect agreement; AC1 better reflects the underlying reliability in those cases.

Table~\ref{tab:icr} summarizes reliability statistics for each code.

\begin{table*}[t]
\centering
\caption[]{Intercoder reliability for the 10 qualitative codes (\textit{N} = 1{,}482 posts).}
\label{tab:icr}

\begin{tabularx}{0.90\textwidth}{>{\raggedright\arraybackslash}X r r r r r r r r}
\toprule
Variable & \% agree & AC1 & $\alpha$ & $\kappa$ & F1 (pos.) & Pos.\ agr. & Disag. & \% zero (C1/C2) \\
\midrule
Personal relevance              & 91.4 & 0.90 & 0.34 & 0.35 & 0.39 & 0.39 & 127 & 90.4 / 95.7 \\
Parasocial bonding              & 88.7 & 0.84 & 0.63 & 0.63 & 0.70 & 0.70 & 167 & 78.1 / 84.5 \\
Productivity boost / automation & 98.0 & 0.98 & 0.67 & 0.67 & 0.68 & 0.68 &  30 & 96.3 / 97.4 \\
Over-reliance decision making   & 99.9 & 1.00 & 0.00 & 0.00 & 0.00 & 0.00 &   1 & 99.9 / 100.0 \\
Private life sharing            & 98.7 & 0.99 & 0.45 & 0.45 & 0.46 & 0.46 &  19 & 98.7 / 98.9 \\
Companionship attribution       & 89.8 & 0.85 & 0.68 & 0.68 & 0.74 & 0.74 & 151 & 77.5 / 83.3 \\
Social isolation mitigation     & 97.5 & 0.97 & 0.82 & 0.82 & 0.84 & 0.84 &  37 & 92.6 / 92.1 \\
Social life enhancement         & 99.8 & 1.00 & 0.40 & 0.40 & 0.40 & 0.40 &   3 & 99.9 / 99.8 \\
Reverse tool perception         & 91.8 & 0.85 & 0.81 & 0.81 & 0.87 & 0.87 & 122 & 67.0 / 67.3 \\
Interpersonal substitution      & 93.6 & 0.93 & 0.29 & 0.31 & 0.33 & 0.33 &  95 & 98.2 / 92.3 \\
\bottomrule
\end{tabularx}

\vspace{2pt}
\begin{minipage}{0.88\textwidth}
\footnotesize\textit{Note.} Codes are non-exclusive binary indicators. Gwet’s AC1 is reported as the primary reliability statistic because several codes are rare and class-imbalanced.
\end{minipage}

\end{table*}

\section{Descriptive Frequencies of Qualitative Codes}
\label{app:sectionD}
For transparency, Table~\ref{tab:keep40_codes} reports descriptive frequencies of our ten qualitative codes based on consensus coding by two coders ($N = 1,482$ posts). A post is counted for a given code only when both coders agreed to assign it. Under this definition, 611 posts (41.2\%) received at least one of the ten codes.

\begin{table*}[t]
\centering
\caption[]{Prevalence of qualitative codes in the \#Keep4o dataset.}
\label{tab:keep40_codes}
\begin{tabular}{lcc}
\toprule
\textbf{Code} & \textbf{Posts with code (n)} & \textbf{\% of all posts (\textit{N} = 1,482)} \\
\midrule
Personal Relevance & 143 & 9.6\% \\
Parasocial Bonding & 325 & 21.9\% \\
Productivity Boost Task Automation & 55 & 3.7\% \\
Over-reliance Decision Making & 1 & 0.1\% \\
Private Life Sharing & 19 & 1.3\% \\
Companionship Attribution & 334 & 22.5\% \\
Social Isolation Mitigation & 110 & 7.4\% \\
Social Life Enhancement & 2 & 0.1\% \\
Reverse Tool Perception & 489 & 33.0\% \\
Interpersonal Substitution Effect & 27 & 1.8\% \\
\bottomrule
\end{tabular}
\end{table*}

\section{Exploratory Process Markers}
\label{app:sectionE}
This section provides additional detail for the exploratory process markers introduced in Section~\ref{sec:variable_operationalization}.

\subsection{Measurement and Coding}
\label{sec:sectionE1}
We explored three linguistically defined \emph{process markers} to probe a potential escalation pathway, treating them as pre-specified \emph{exploratory} variables:
\begin{itemize}
  \item \textbf{Temporal}: sequential phrasing (e.g., ``after/since/when [removed/changed]'').
  \item \textbf{Causal}: explicit causation (e.g., ``because/led to/resulted in'').
  \item \textbf{Escalation}: forced adaptation (e.g., ``now I have to \ldots'').
\end{itemize}

Markers were coded via case-insensitive string matching with basic lemmatization; posts could match multiple categories. We maintained strict separation between exposure and outcome lexicons to avoid circularity. Exposure used the strict threshold of scores $\ge 2$ (explicit choice removal or coercive framings) versus $<2$; sample sizes were $n=72$ (strict) and $n=1{,}410$ (non-strict). See Section~\ref{sec:quantitative_method} for variable definitions and coding details.

\subsection{Statistical Reporting}
\label{sec:sectionE2}
For each marker, we report counts/rates, absolute differences (percentage points), and 95\% CIs. When sparse data produced zeros, we provide 95\% upper bounds using the rule-of-three ($\approx 3/n$). For comparative effect sizes, we report risk ratios (RR) with 95\% CIs using the Katz log method with Haldane--Anscombe correction where needed. Given the small $n$ in the strict group, we emphasize effect sizes and transparency of counts rather than significance testing (see Table~\ref{tab:process_markers}).

\subsection{Results}
\label{sec:sectionE3}
Table~\ref{tab:process_markers} summarises the exploratory process-marker results.
\begin{table*}[t]
\centering
\caption[]{Process markers by exposure group (exploratory).}
\label{tab:process_markers}

\begin{tabular}{@{}lccc@{}}
\toprule
\textbf{Marker} & \textbf{Strict ($n=72$)} & \textbf{Non-strict ($n=1{,}410$)} & \textbf{Note} \\
\midrule
Temporal   & 0/72 ($\leq 4.2\%$)                 & 0/1410 ($\leq 0.21\%$)            & not observed \\
Causal     & 6/72 (8.3\%; 95\% CI 3.9--17.0)     & 72/1410 (5.1\%; 95\% CI 4.1--6.4) & RR=1.63 [0.73, 3.63]; $\Delta$=+3.2 pp \\
Escalation & 0/72 ($\leq 4.2\%$)                 & 0/1410 ($\leq 0.21\%$)            & near-zero under conservative lexicon \\
\bottomrule
\end{tabular}

\vspace{2pt}
\begin{minipage}{0.82\textwidth}
\footnotesize\textit{Note.} Strict exposure denotes scores $\geq 2$; non-strict denotes scores $<2$.
For zero counts, we report 95\% upper bounds via the rule-of-three ($\approx 3/n$).
\end{minipage}
\end{table*}

\subsection{Interpretation and Limits}
\label{sec:sectionE4}
Table~\ref{tab:process_markers} shows a modest uptick in explicit causal connectors under strict choice-deprivation (6/72, 8.3\%) versus other posts (72/1410, 5.1\%; RR=1.63, 95\% CI [0.73, 3.63]). By contrast, temporal and escalation templates were not observed (95\% upper bounds $\leq 4.2\%$ and $\leq 0.21\%$), likely reflecting short-form posts and our conservative lexicon. Given the sparse counts and wide intervals, we treat these markers as suggestive but limited support rather than confirmation of the mechanism. They complement---rather than replace---the main evidence that choice deprivation is selectively and non-linearly associated with rights/procedural protest (Panels A/B in Table~\ref{tab:chi42}).

\section{Human Coding Reliability and Lexicon Validation (Quantitative Variables)}
\label{app:sectionF}

\subsection{Construct Overview and Example Phrases}
\label{sec:sectionF1}
Table~\ref{tab:constructs} summarizes the constructs used in Section~\ref{sec:quantitative_method} and the lexicon-based implementation, with short example phrases (non-exhaustive). The full pattern lists appear in our analysis code repository.

\begin{table*}[t]
\centering
\caption[]{Construct classification framework}
\Description{Construct classification framework}
\label{tab:constructs}
\begingroup
\setlength{\emergencystretch}{3em}
\begin{tabularx}{\textwidth}{ll>{\raggedright\arraybackslash}X>{\raggedright\arraybackslash}X}
\toprule
\multicolumn{1}{c}{\textbf{Construct}} &
\multicolumn{1}{c}{\textbf{Type/tier}} &
\multicolumn{1}{c}{\textbf{Intended meaning}} &
\multicolumn{1}{c}{\textbf{Example phrases (not exhaustive)}} \\
\midrule
Choice-deprivation 1 & Implicit disruption & Sudden, non-consensual change without explicit ``no choice'' language & ``button is gone'', ``suddenly changed'', ``without warning'' \\
Choice-deprivation 2 & Explicit removal of choice & Directly stating that users cannot select or opt out & ``no option'', ``can't choose'', ``decision made for us'', ``selector/toggle/legacy'' \\
Choice-deprivation 3 & Coercive framing & Language of force, imposition, or non-consent & ``forced'', ``yanked away'', ``removed without consent'', ``dictated'', ``powerless'' \\
Protest-Rights & Outcome & Rights, democracy, consent, agency & ``rights'', ``choice'', ``democracy'', ``voice'', ``user control'', ``consent'' \\
Protest-Relational & Outcome & Grief, betrayal, attachment to the model & ``betrayal'', ``grief'', ``goodbye'', ``friend'', ``soul'', ``trauma'', ``trust'' \\
Process-Temporal & Process marker & Before/after/since the change & ``after they removed\ldots'', ``since the switch\ldots'', ``when you changed\ldots'' \\
Process-Causal & Process marker & Explicit because/led-to language & ``because\ldots'', ``led to\ldots'', ``resulted in\ldots'', ``therefore\ldots'' \\
Process-Escalation & Process marker & Added work or workarounds (e.g., ``Now I have to\ldots'') & ``now I have to\ldots'', ``this forces me to\ldots'', ``I'm forced to\ldots'' \\
\bottomrule
\end{tabularx}
\endgroup
\end{table*}

\subsection{Intercoder Reliability for Quantitative Codes}
\label{sec:sectionF2}
Two coders independently annotated all 1,482 posts for (i) the 4-level choice-deprivation intensity scale (0--3), (ii) the presence of the Protest-Rights frame (binary), and (iii) a 4-category process marker (0 = none, A = temporal, B = causal, C = escalation). We treat these annotations as quantitative codes used to derive the exposure and outcome variables in Section~\ref{sec:quantitative_method}.

Table~\ref{tab:intercoderquantitative} summarizes intercoder agreement for the three variables. The deprivation scale and Protest-Rights frame show substantial agreement, while the four-way process label reaches moderate agreement given its subtle distinctions and low base rates for A/B/C.

For the deprivation scale, the $4\times 4$ confusion matrix (not shown) indicates that most disagreements involve adjacent categories (e.g., 1 vs.\ 2 rather than 0 vs.\ 3), consistent with the ordinal and fuzzy nature of the construct. In the main text, we therefore collapse this scale into binary thresholds ($\geq 1$ vs.\ 0; $\geq 2$ vs.\ $<2$) to reduce boundary noise while retaining its coarse severity information.

\begin{table*}[t]
\centering
\caption[]{Intercoder reliability for quantitative codes (coder1 vs coder2, \textit{N} = 1,482 posts).}
\label{tab:intercoderquantitative}

\begin{tabularx}{\textwidth}{>{\raggedright\arraybackslash}X >{\raggedright\arraybackslash}X c c c c}
\toprule
\textbf{Variable} & \textbf{Coding scheme} & \textbf{\textit{N}} & \textbf{Percent agreement} & \textbf{Cohen's $\kappa$} & \textbf{Weighted $\kappa$ (quadratic)} \\
\midrule
Choice-Deprivation intensity & 0--3 ordinal & 1482 & 0.791 & 0.669 & 0.879 \\
Protest-Rights frame & 0/1 (absent / present) & 1482 & 0.919 & 0.722 & -- \\
Process label (temporal/causal/escalation) & 0 / A / B / C (categorical) & 1482 & 0.810 & 0.566 & -- \\
\bottomrule
\end{tabularx}

\end{table*}

\subsection{Lexicon validation against reconciled human codes}
\label{sec:sectionF3}
After computing intercoder reliability for the quantitative codes (Appendix~\ref{sec:sectionF2}), the two coders met to reconcile disagreements for the choice-deprivation and Protest-Rights variables. Through discussion and, where needed, adjudication by the first author, they produced a single reconciled label per post; we refer to these as the \texttt{last\_} codes (e.g., \texttt{last\_Choice\_Deprivation\_Grade},\newline \texttt{last\_Protest\_Rights\_Score}). These reconciled codes serve as our human ``consensus'' reference for validating the lexicon-based measures. For the exploratory Process-Causal marker, we retain coder1's label (``B'' vs.\ 0/A/C) as the reference given moderate intercoder reliability ($\kappa \approx 0.57$; Appendix~\ref{sec:sectionF2}) and the marker's secondary role in the analysis.

To evaluate how well the lexicon-based measures approximate these human judgments, we compared lexicon outputs to the reconciled human codes for three constructs: choice-deprivation intensity (under two binary thresholds), the Protest-Rights frame, and the Process-Causal marker. For choice deprivation, we used the reconciled 4-level score (\texttt{last\_Choice\_Deprivation\_Grade}, 0--3) and binarized it at $\geq 1$ (``any deprivation'') and $\geq 2$ (``strict deprivation''). For Protest-Rights, we compared the lexicon prediction to the reconciled binary code (\texttt{last\_Protest\_Rights\_Score}). For the Process-Causal marker, we treated coder1's label ``B'' (causal) versus all other labels (0/A/C) as the positive class and compared it to the \texttt{Process\_Causal\_lex} flag. All comparisons were computed after deduplicating posts by \texttt{Post\_ID}, yielding an analytical corpus of 1,482 unique posts.

Table~\ref{tab:lexicon-human} reports percent agreement, Cohen's $\kappa$, and positive-class precision, recall, and F1 for each construct, treating the reconciled human labels as the reference.

\begin{table*}[t]
\centering
\caption[]{Comparison of lexicon-based measures to human codes.}
\label{tab:lexicon-human}

\begin{tabularx}{\textwidth}{Xcccccc}
\toprule
\textbf{Construct (positive class)} & \textbf{\textit{N}} & \textbf{Percent agreement} & \textbf{Cohen's $\kappa$} & \textbf{Precision} & \textbf{Recall} & \textbf{F1} \\
\midrule
Choice-Deprivation (ANY, score $\geq 1$ vs.\ 0) & 1482 & 0.590 & 0.097 & 0.843 & 0.105 & 0.187 \\
Choice-Deprivation (STRICT, score $\geq 2$ vs.\ $<2$) & 1482 & 0.778 & 0.134 & 0.583 & 0.123 & 0.203 \\
Protest-Rights (rights frame present) & 1482 & 0.877 & 0.520 & 0.573 & 0.613 & 0.592 \\
Process-Causal (B vs.\ non-B; causal marker present) & 1482 & 0.842 & 0.165 & 0.449 & 0.155 & 0.230 \\
\bottomrule
\end{tabularx}

\end{table*}

The Protest-Rights lexicon attains medium agreement with the reconciled human codes ($\kappa \approx 0.52$) and reasonably balanced precision and recall ($\approx 0.57$--$0.61$), indicating that it captures many clear instances of rights-based protest language. By contrast, the choice-deprivation and Process-Causal lexicons behave as conservative indicators: they show relatively high precision ($\approx 0.58$--$0.84$) but low recall ($\approx 0.10$--$0.16$), flagging only a subset of ``on-the-nose'' explicit formulations (e.g., ``no option,'' ``forced,'' ``because/led to/resulted in'') while missing more implicit or creatively phrased cases. We therefore treat these lexicon-based variables as coarse but interpretable measurement tools rather than high-accuracy classifiers, and use them in the main analysis (Section~\ref{sec:section4.2}) to identify broad associative patterns that complement---rather than replace---the richer, human-coded qualitative findings.

\end{document}